\documentclass[a4paper,11pt]{article}
\usepackage{pos}

\title{Hadron-Hadron Interactions from Lattice QCD: \\ Theory meets Experiments}

\author*{Tetsuo Hatsuda}

\affiliation{RIKEN Center for Interdisciplinary Theoretical and Mathematical Sciences (iTHEMS),\\
 RIKEN,  Wako, Saitama 351-0198, Japan}

\affiliation{Kavli Institute for the Physics and Mathematics of the Universe (Kavli IPMU), WPI,  \\
The University of Tokyo, Kashiwa, Chiba 277-8568, Japan}

\emailAdd{thatsuda@riken.jp}

\abstract{

We summarize recent developments in the study of hadron-hadron interactions using lattice QCD near the physical pion mass ($m_{\pi} \simeq 146$~MeV), based on the HAL QCD method and its connection to experimental data. In particular, we focus on several key interaction channels shown below.
Also,  we examine the two-pion exchange (TPE) mechanism, which governs the long-range behavior of interactions between flavor-singlet hadrons, as well as between nucleons and flavor-singlet hadrons.

\vspace{2em}

\hspace{1em} \text{\bf Table of Contents}
   
\begin{itemize}
    \item[1.] Introduction
    \item[2.] The HAL QCD Method
    \item[3.] Baryon-Baryon Interactions: $\Lambda$-$\Lambda$ and $\Xi$-$N$
    \item[4.] Meson-Meson Interactions: $D^*$-$D$ and $T_{\rm cc}$
    \item[5.] Meson-Baryon Interactions with strangeness: $\phi$-$N$
    \item[6.] Meson-Baryon Interactions with charm: $J/\psi$-$N$ and $\eta_c$-$N$
    \item[7.] Two-pion Exchange (TPE) Paradigm
        \begin{itemize}
        \item[7.1] TPE in nuclear force
        \item[7.2] TPE between flavor-singlet hadrons
        \item[7.3] TPE between nucleon and flavor-singlet hadron
        \item[7.4] Signatures of TPE in $\phi$-N, $J/\psi$-N and $\eta_c$-N interactions
    \end{itemize}
   \item[8.] Summary
\end{itemize}
}

\FullConference{The XVIth Quark Confinement and the Hadron Spectrum Conference (QCHSC24)\\
 19-24 August, 2024\\
 Cairns Convention Centre, Cairns, Queensland, Australia\\}

\begin{document}
\maketitle

\section{Introduction}

\begin{figure*}[b]
    \centering
   \includegraphics[width=10.0cm]{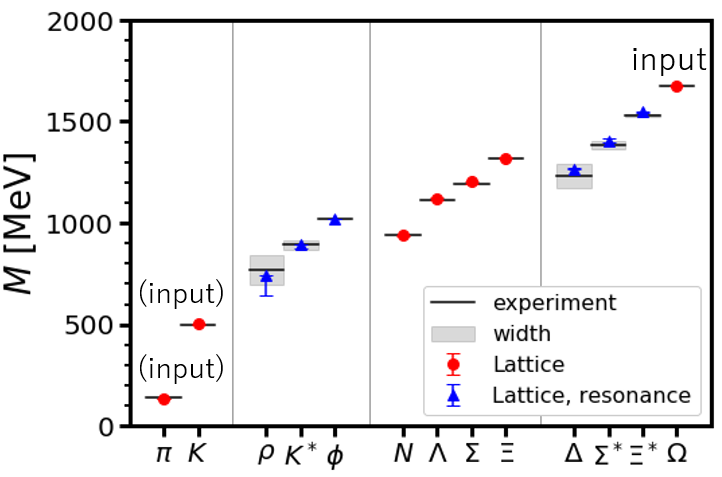}
   \caption{Hadron spectrum at the physical point \((m_{\pi} = 137\) MeV) from (2+1)-flavor lattice QCD simulations~\cite{Fugaku-conf}. Red circles denote the masses of stable hadrons, while blue triangles represent the masses of hadronic resonances, with statistical and systematic uncertainties combined in quadrature. Horizontal black lines indicate the experimental values, and gray bands highlight the decay widths of the resonances.
}
    \label{Fig.1}
\end{figure*}

Quantum chromodynamics (QCD) underlies the properties of hadrons, atomic nuclei, and even the internal structure of neutron stars. Currently, lattice QCD stands as the only first-principles approach capable of revealing the various nonperturbative characteristics of QCD through large-scale Monte Carlo simulations. Decades of theoretical advancements, coupled with the advent of powerful supercomputers, have enabled the accurate reproduction of fundamental properties of individual hadrons.  For example, Fig.\ref{Fig.1} presents the hadron masses obtained from isospin-symmetric simulations conducted by the HAL QCD collaboration using the 440 PFlops Fugaku supercomputer at RIKEN. These simulations were performed at the physical point, corresponding to a pion mass of \( m_\pi \simeq 137 \) MeV, within (2+1)-flavor QCD on a hypercubic lattice lattice with the lattice spacing of \( a = 0.084 \) fm and a large volume of \( V = (8.1 \, {\rm fm})^3 \) ~\cite{Fugaku-conf}.

The time is now ripe to extend these quantitative calculations to multi-hadron systems, such as baryon-baryon, meson-meson, and baryon-meson interactions, using lattice QCD simulations near and at the physical pion mass. Such precise numerical data are invaluable for understanding the structure of ordinary nuclei and hypernuclei, as well as for exploring the equation of state of high-density matter~\cite{Baym2018}. Moreover, these interactions serve as fundamental inputs for studying exotic hadrons and mesic nuclei, offering critical insights into the realization of confinement and chiral symmetry breaking, both in the vacuum and within hadronic matter~\cite{Hayano2008}.

There are two primary approaches in lattice QCD for studying hadron-hadron interactions. The first is the HAL QCD method~\cite{HAL1,HAL2,HAL3,HAL4}, which determines an energy-independent, non-local potential by analyzing the spatiotemporal correlations between two hadrons. This potential, derived from the Haag-Nishijima-Zimmermann (HNZ) reduction formula for the four-point correlation function of composite operators (see~\cite{Reduction} and references therein), enables the calculation of binding energies and phase shifts in infinite volume through a Lippmann-Schwinger type integro-differential equation formulated from the correlation function.
The second approach is the direct method, which extracts binding energies and scattering phase shifts from lattice eigenenergies using L\"{u}scher’s finite-volume formula~\cite{FVM1986}, derived from the temporal correlations between two hadrons.
  
One of the key challenges in studying two-hadron systems, compared to single hadrons, is the presence of elastic scattering states. The corresponding excitation energies, \( \Delta E \), are typically one to two orders of magnitude smaller than \( O(\Lambda_{\text{QCD}}) \). Consequently, the direct method requires probing large Euclidean times \( t \gtrsim (\Delta E)^{-1} \) to achieve ground-state saturation and extract the true signal. However, statistical uncertainties grow exponentially with both Euclidean time \( t \) and the number of baryons \( A \), making it practically impossible to reliably isolate the true ground state using this approach.
In contrast, the HAL QCD method circumvents the ground-state saturation problem entirely. By extracting an energy-independent potential from the spatial and temporal correlations of hadrons, it accounts for contributions from both ground and excited elastic scattering states. For further details comparing the two approaches, see Ref.~\cite{HALReview} and references therein.

In the present article,  some hadron-hadron interactions
calculated in (2+1)-flavor lattice QCD 
with  \( a = 0.0846 \) fm and a large volume of \( V = (8.12 \, {\rm fm})^3 \) 
and with the pion mass  $m_\pi = 146.4$ MeV are
 presented together with its comparison to the experimental data.
 Underlying gauge configurations near the physical pion mass were
 carried out by the 11 PFlops K computer at RIKEN \cite{K-conf}, and 
is called the ``K-configuration" below.

\section{The HAL QCD Method}

The starting point of the HAL QCD method 
is the 4-point correlation function ($F(\vec{r}, t)$)
of two hadrons ($h_1$ and $h_2$):
\begin{equation}
F(\vec{r}, t) \equiv \langle 0 | T \left[ \sum_{\vec{x}} h_1(\vec{x} + \vec{r}, t) h_2(\vec{x}, t) \mathcal{J}_{h_1h_2}^{\mathrm{src}}(0) \right] | 0 \rangle
= \sum_n A_n \psi_{n}(\vec{r}) e^{-E_n t} + \cdots,
\end{equation}
where $h_{1}$ and $h_2$ are the hadronic interpolating operators or the 
sink operators at time $t > 0$  and  $\mathcal{J}_{h_1h_2}^{\mathrm{src}}$ is a source  operator 
 at $t=0$.  Also,  $A_n \equiv \langle h_1+h_2, E_n | \mathcal{J}_{h_1 h_2}^{\mathrm{src}}(0) | 0 \rangle$ represents the overlap factor, while the ellipsis denotes contributions from inelastic states. $\psi_{n}(\vec{r})$ is the equal-time Nambu-Bethe-Salpeter amplitude or the  NBS wave function in short.
  The so-called ``$\mathcal{R}$-correlator'' is defined as
\begin{equation}
\mathcal{R}(\vec{r}, t) \equiv \frac{F(\vec{r}, t)}{C_{h_1}(t)C_{h_2}(t)}
= \sum_n \frac{A_n}{Z_{h_1}Z_{h_2}} \psi_{n}(\vec{r}) e^{-(E_n - (m_{h_1} +m_{h_2}) t} + \cdots, 
\end{equation}
where $C_{h_1}(t)$ and $C_{h_2}(t)$ are the 2-point hadronic correlation functions.
The elastic component of $\mathcal{R}(\vec{r}, t)$ satisfies the following integro-differential equation,
\begin{equation}
\label{eq:NBS-eq}
\left[ \frac{1 + 3\delta^2}{8\mu} \frac{\partial^2}{\partial t^2} - \frac{\partial}{\partial t} - H_0 + O(\delta^2 \partial_t^3) \right] \mathcal{R}(\vec{r}, t)
= \int d^3r' U(\vec{r}, \vec{r}') \mathcal{R}(\vec{r}', t),
\end{equation}
with $\mu = (m_{h_1} m_{h_2})/(m_{h_1} + m_{h_2})$ and 
$\delta = ({m_{h_1} - m_{h_2}})/({m_{h_1} + m_{h_2}})$.  In the examples shown in this article, 
 the higher derivative terms of $O(\delta^2 \partial_t^3)$ are small and negligible.

Below the inelastic threshold $E_{\mathrm{th}}$, the potential $U(\vec{r}, \vec{r'})$ accurately reflects the phase shifts, as encoded in the asymptotic behavior of the NBS wave function at large distances. Eq.(\ref{eq:NBS-eq})  neither requires ground-state saturation nor the determination of individual eigenenergies $E_n$. Thus, unlike the direct method, the conditions for reliable calculations in the  HAL QCD method are significantly less stringent, requiring only $t \gtrsim O(\Lambda_{\mathrm{QCD}}^{-1}) \sim O(1)$ fm.

In most of the practical applications at low energies, the non-local potential can be  expanded as
\begin{equation}
U(\vec{r}, \vec{r'}) = \sum_n V_n(\vec{r}) \nabla^n \delta(\vec{r} - \vec{r'}).
\end{equation}
In particular, for the spin-singlet channel, the leading-order (LO) truncation of the expansion is given by
\begin{equation}
\label{eq:V0} 
V_0^{\rm LO}(r) = \mathcal{R}^{-1}(r, t)
\left[ \frac{1 + 3\delta^2}{8\mu} \frac{\partial^2}{\partial t^2} - \frac{\partial}{\partial t} - H_0 \right]
\mathcal{R}(r, t).
\end{equation}
One can estimate the systematic error of the truncation of the derivative expansion 
 by studying the $t$ dependence of the right hand side of Eq.(\ref{eq:V0}). 
 
Instead of using the derivative expansion, one may also invert Eq.(\ref{eq:NBS-eq})
 directly to extract the non-local potential $U(\vec{r}, \vec{r'})$ 
 by using the deep neural network with the loss function, ${\cal L} = \sum_{\vec{r}, t} |({\rm LHS})-({\rm RHS})|^2$
  with appropriate regularization,
  where LHS (RHS) is the left hand side (right hand side)  of  Eq.(\ref{eq:NBS-eq})  \cite{Wang2025}.

\section{Baryon-Baryon Interaction: $\Lambda$-$\Lambda$ and $\Xi$-$N$} 

\begin{figure*}[b]
    \centering
    \includegraphics[width=7.0cm]{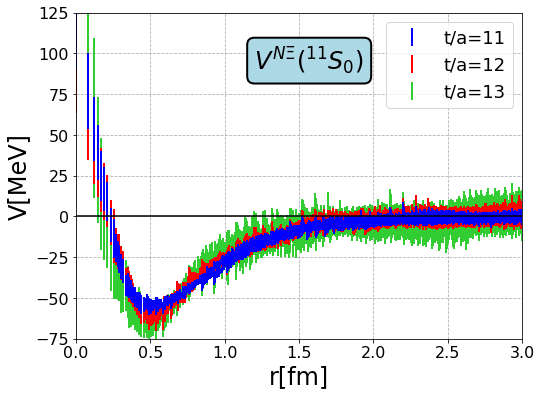}\hspace{0.5cm}
   \includegraphics[width=6.8cm]{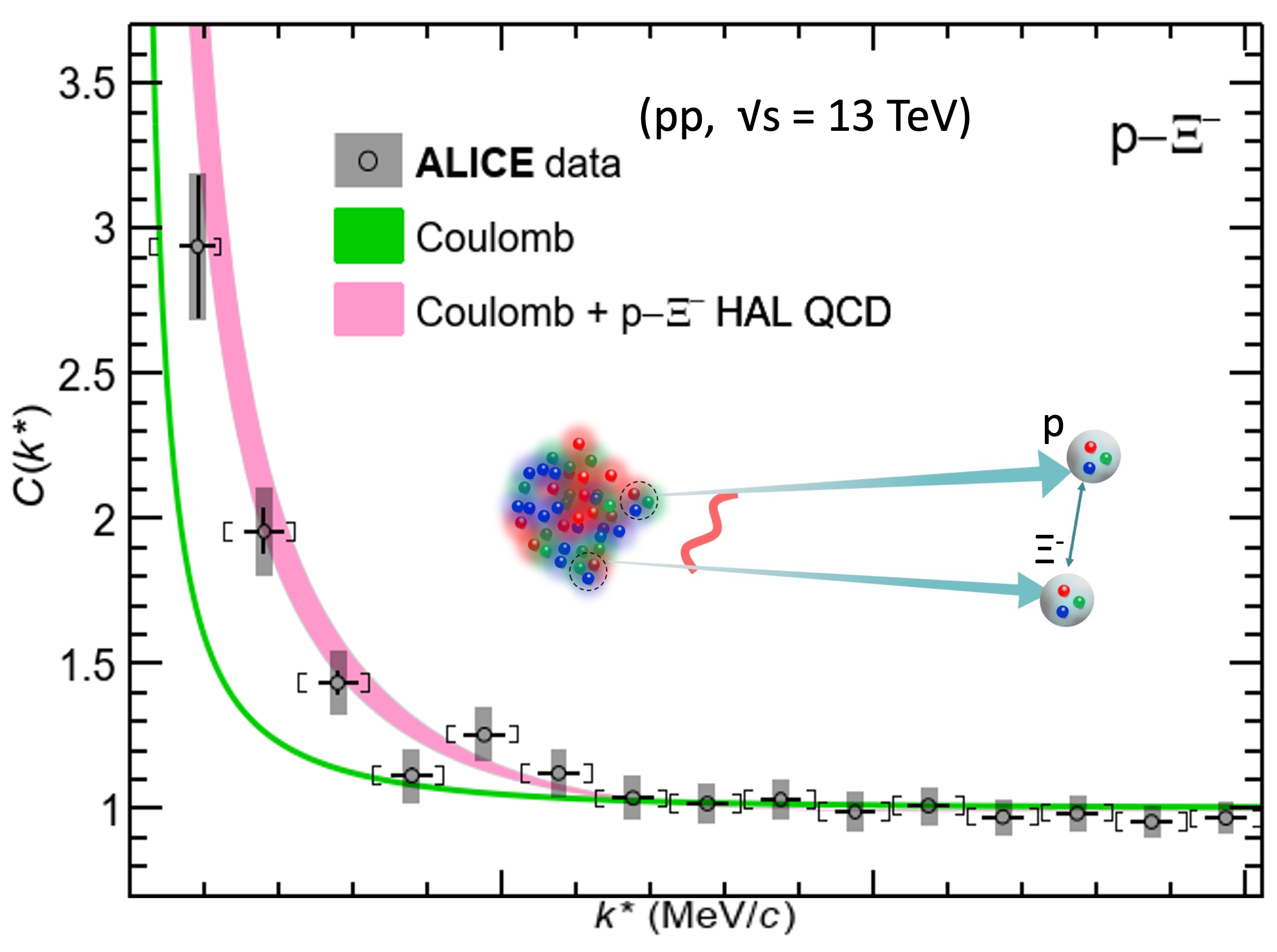} 
   \caption{(Left) The diagonal part of the $\Xi N$ potential in the spin-isospin singlet channel \cite{Sasaki2020}.
  (Right) Femtoscopic $\Xi^-$-proton correlation measured by ALICE collaboration at the LHC $pp$ collisions  \cite{ALICE2020}. The green curve 
  corresponds to the case of Coulomb attraction only, while
  the pink band includes the extra attraction by the stong interaction calculated by HAL QCD Collaboration \cite{Sasaki2020}.}
    \label{Fig.2}
\end{figure*} 

Let us start with the  hyperon interaction with strangeless $=-2$ as one of the characteristic examples to show the predictive power of lattice QCD. Using the  HAL QCD method, $\Lambda\Lambda$- $\Xi N$ coupled-channel lattice QCD potential has been extracted for four distinct spin-isospin channels and parametrized by some analytic functions to determine the observables such as the scattering phase shifts \cite{Sasaki2020}. It was found that
  the transition potential between $\Lambda\Lambda$ and $\Xi N$ is found to be significant only at
  short distance and does not affect the 
   phase shift  except near the 
   $\Xi N$ threshold in the $\Lambda\Lambda$ scattering.  Also, 
    the weak $\Lambda\Lambda$ attraction is unlikely to support a bound dihyperon (the $H$ dibaryon)  below the $\Lambda\Lambda$ threshold.
   On the other hand, the $\Xi N$ interaction in the spin-isospin singlet channel ($^{11}S_0$) proves to be the most attractive
   as shown in Fig.\ref{Fig.2} (left panel), bringing the $\Xi N$ system close to unitarity. This finding has implication
   to the final state interaction of the $\Xi N$ pair  produced in 
   high-energy collision experiments (see \cite{Kamiya2022} and references therein), as well
    as the structure of 
    hypernuclei with strangeness=-2 (see \cite{Isaka2024} and references therein).

   Shown in  Fig.\ref{Fig.2} (right panel)
    is an experimental result of the femtoscopic correlation $C(k^*)$ between  $\Xi^-$ and the proton
       obtained by the ALICE Collaboration at LHC \cite{ALICE2020}:
    \begin{equation}
  C(k^*) = \int d^3 r \sum_j \omega_j S_j(\vec{r}) \left| \Psi_j^{(-)}(\vec{r};k^*) \right|^2.
    \end{equation}
Here, $k^*$ denotes the relative momentum in the center-of-mass frame of the particle pair. The wave function \(\Psi_j^{(-)}\) in the \(j\)th channel is expressed as a function of the relative coordinate \(\vec{r}\) in that channel, with an outgoing boundary condition applied in the measured channel. 
The functions \(S_j(\vec{r})\) and \(\omega_j\) represent the normalized source function and its corresponding weight in the \(j\)th channel, respectively (see \cite{Kamiya2022} for details).
In Fig.~\ref{Fig.2} (right panel), the experimental data show an additional $\Xi^- p$ attraction beyond the Coulomb interaction at low $k^*$. This difference can be explained {\it quantitatively} by the attractive interaction predicted by lattice QCD results, without any fine-tuning.

This analysis marks the first quantitative comparison between lattice QCD predictions for hadron-hadron interactions and experimental data from high-energy collisions.
\section{Meson-Meson Interactions: $D^*$-$D$ and $T_{\rm cc}$}

\begin{figure*}[b]
    \centering
    \includegraphics[width=7.3cm]{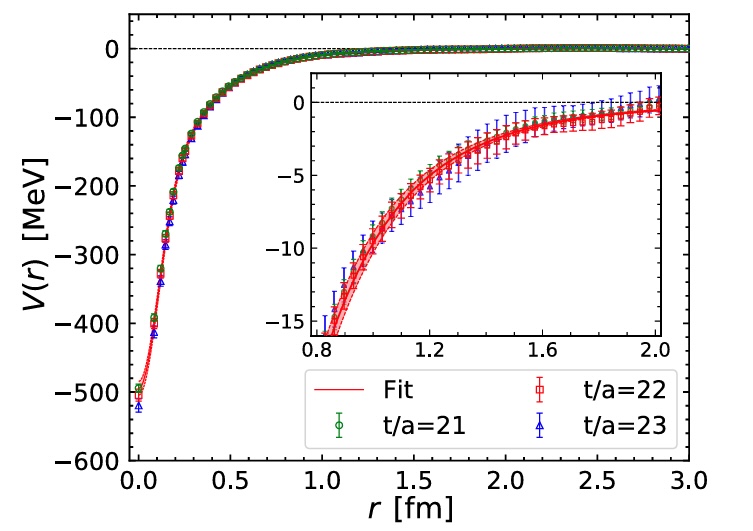}\hspace{0.5cm}
   \includegraphics[width=7.0cm]{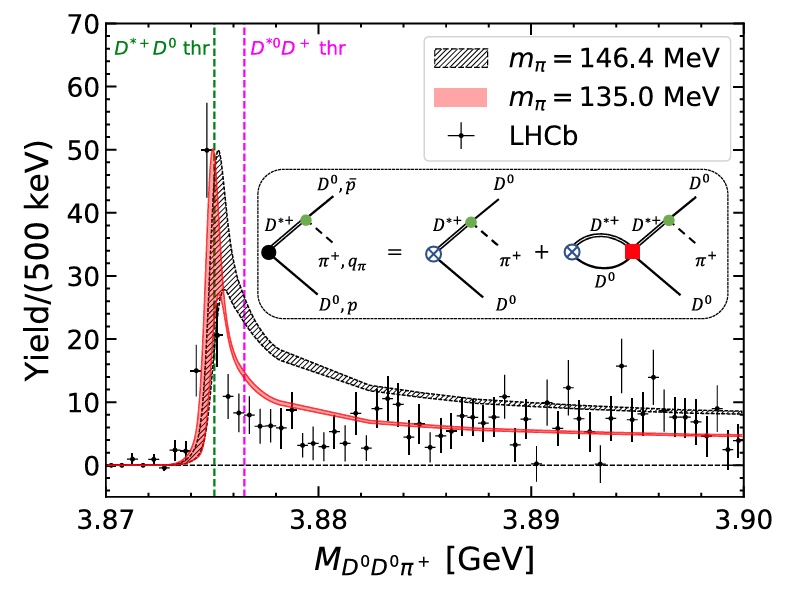} 
   \caption{(Left) The $D^*D$ potential $V(r)$ in the $I=0$ and $S$-wave channel at Euclidean time $t/a = 21, 22, 23$. The red band shows the fitted potential with $V_{\text{fit}}$ in
    Eq.(\ref{eq:fit}) for $t/a = 22$. The inset shows a magnification.
(Right) The $D^0 D^0 \pi^+$ mass spectrum. Theoretical results with $V_{\text{fit}} (r; m_\pi)$ in Eq.(\ref{eq:fit}) for $m_\pi = 146.4$ MeV ($m_\pi = 135.0$ MeV) are shown by the black (red) band. The black points are LHCb data~\cite{LHCb2022}. The inset shows diagrams contributing to the $D^0 D^0 \pi^+$ mass spectrum, where the black filled circle, blue cross circle, green filled circle, and red square denote production amplitude $U$, constant vertex $P$, $D^{*+} \to D^0 \pi^+$ vertex, and scattering $T$ matrix, respectively.}
    \label{Fig.3}
\end{figure*} 

The search for exotic hadrons with multi-quark configurations beyond the conventional constituent quark model has been a central topic in the study of nonperturbative QCD for decades. While numerous candidates for exotic hadrons have been reported, the LHCb Collaboration recently identified a doubly charmed tetraquark \(T_{cc}^+\)  \cite{LHCb2022}:
A distinct narrow peak is observed in the \(D^0D^0\pi^+\) mass spectrum, located approximately 360 keV below the \(D^{*+}D^0\) threshold. The isospin \(I\), spin \(J\), and parity \(P\) of this state are consistent with \((I, J^P) = (0, 1^+)\).
 HAL QCD Collaboration has studied the interaction between \(D^*\) and \(D\) mesons in the isoscalar \(S\)-wave channel, derived through hadronic spacetime correlations
  to unravel the structure of $T_{cc}^+$ \cite{Lyu2024}. They found that
   an attractive nature between $D^*$ and $D$ across all distances as shown in the left panel of Fig.\ref{Fig.3}.
    Here the following fit function (two-range Gaussian potential at short distance together with the  two-pion exchange potential  at intermediate and long distances) was used 
    to draw the red band in the figure:
\begin{equation}
\label{eq:fit}
V_{\text{fit}}(r; m_\pi) = \sum_{i=1,2} a_i e^{-(r/b_i)^2} + a_3 
\left[ \left(1 - e^{-(r/b_3)^2}\right) V_{\pi}(r) \right]^2,
\end{equation}
with $V_{\pi}(r) = {e^{-m_{\pi} r}}/{r}$.   This interaction generates a near-threshold virtual state at $m_{\pi}=146$ MeV.
As the pion mass decreases toward its physical value, the virtual state gradually transitions into a loosely bound state guided by the pion-exchange interaction. This feature provides a semi-quantitative interpretation of the \(D^0D^0\pi^+\) mass spectrum in the LHCb measurement  as shown in the right panel of Fig.\ref{Fig.3}.

This is the first quantitative comparison between LHCb data and lattice QCD data at the level of the observed mass spectrum.

\section{Meson-Baryon Interactions with strangeness: $\phi$-$N$}

\begin{figure*}[b]
    \centering
    \includegraphics[width=7.0cm]{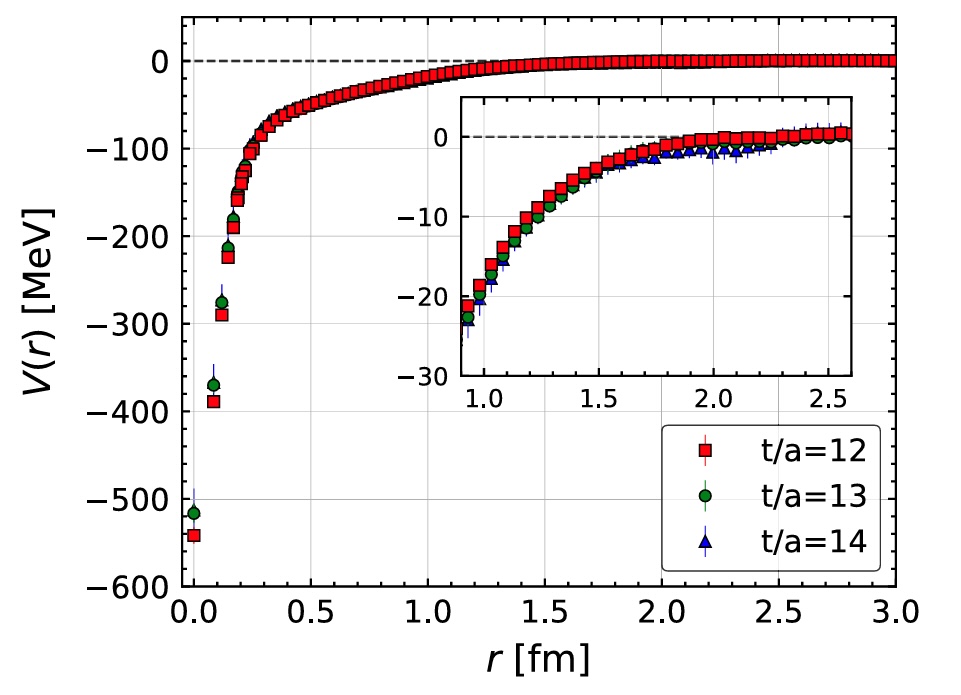}\hspace{0.2cm} 
   \includegraphics[width=6.4cm]{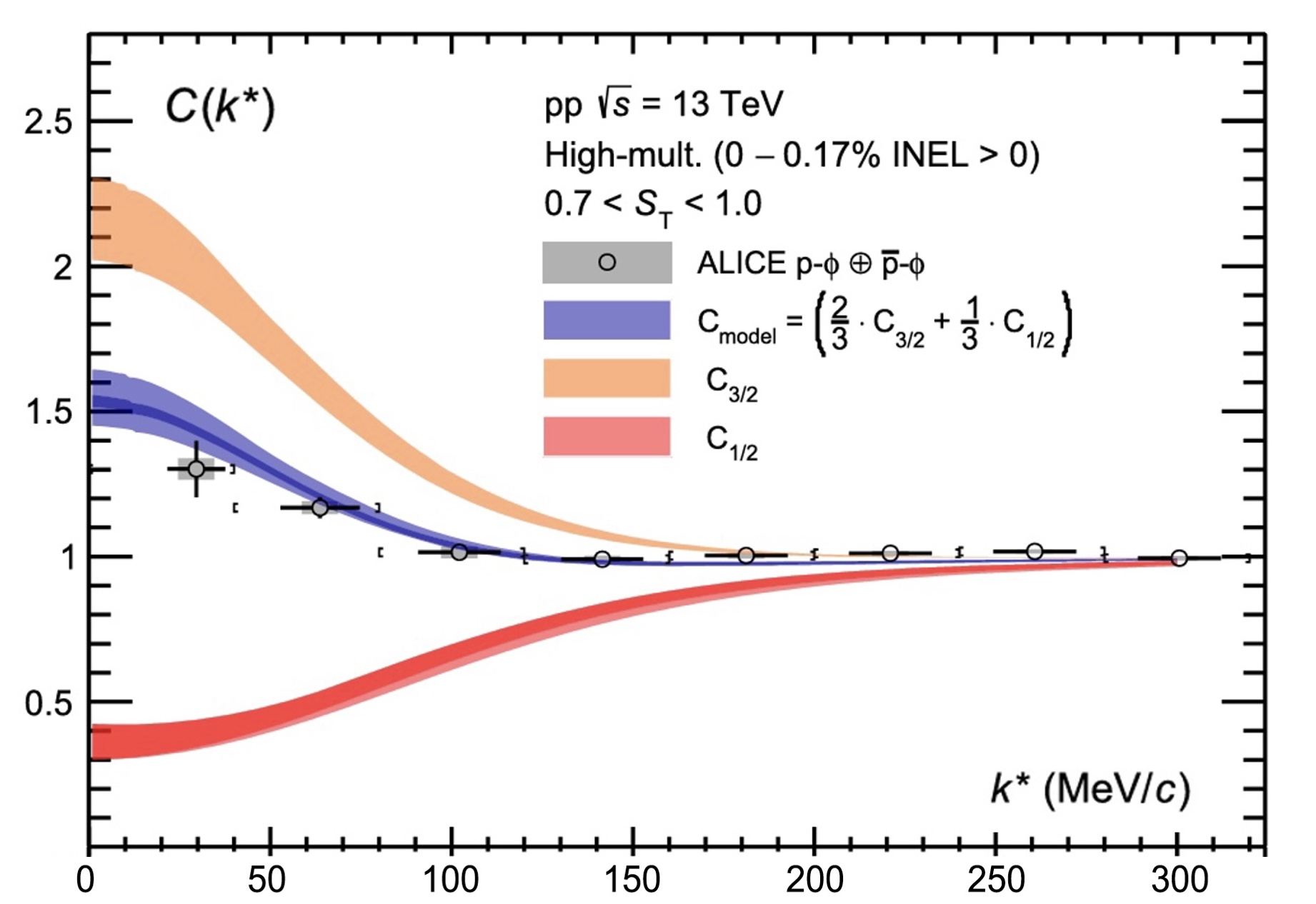} 
   \caption{(Left) The $\phi$-$N$ potential $V(r)$ in the ${}^4S_{3/2}$ channel as a function of separation $r$ at Euclidean time $t/a = 12, 13, 14$ \cite{Lyu:2022imf}. (Right) The experimental
   $\phi$-$p$ correlation function 
   \cite{ALICE2021} together with the spin-averaged model correlation function (blue band) and the unweighted ${}^4S_{3/2}$ (orange band) and ${}^2S_{1/2}$ contributions (red band) \cite{Chizzali2024}.}
    \label{Fig.4}
\end{figure*}

The existence of $\phi$-mesic bound states with nucleons 
remains a long-standing open question. A precise understanding of the $\phi$-$N$ strong interaction serves as the foundation for exploring the possibility of such bound states. 
The ALICE collaboration recently reported a significant attractive interaction between the $\phi$ meson and the proton, based on measurements of the $\phi$-proton correlation function $C(k^*)$ in momentum space from $pp$ collisions at the LHC \cite{ALICE2021}.

  HAL QCD Collaboration presented the first lattice simulation of the $\phi$-$N$ system by focusing on the highest-spin state, specifically the ${}^4S_{3/2}$ channel \cite{Lyu:2022imf}.  
 Here we use the spectroscopic notation ${}^{2s+1}L_J$ ($s$: total spin, $L$: orbital angular momentum, $J$: total angular momentum). This particular channel is of interest because its coupling to two-body open channels, such as $\Lambda K ({}^2D_{3/2})$ and $\Sigma K ({}^2D_{3/2})$, is kinematically suppressed at low energies due to their $D$-wave nature. Furthermore, decay processes involving three or more final-state particles, such as $\Sigma \pi K$, $\Lambda \pi K$, and $\Lambda \pi \pi K$, are also expected to be suppressed due to limited phase space.  The potential $V(r)$ in the ${}^4S_{3/2}$ channel shown in Fig.\ref{Fig.4} (left panel)  is indeed attractive  for all distances and provides the scattering length 
  qualitatively consistent with the value extracted from the experimental data of the ALICE Collaboration. Also, the potential has a characteristic two-component structure, the attractive core at short distance and the attractive tail at long distance. This is similar to the case of the $N\Omega({}^5S_2)$ potential previously studied~\cite{Iritani2019}:
 The Pauli exclusion principle between quarks does not operate in both cases, since $\phi$ and $\Omega$  have no common valence quarks with $N$.

One may go one step further and re-analyze the experimental $\phi$-$p$ correlation function  incorporating constraints from the spin 3/2 channel 
 from lattice QCD \cite{Chizzali2024}. This approach enabled the extraction of the spin 1/2 channel for the first time, employing a phenomenological complex potential inspired by lattice QCD calculations. The real part of the potential is found to be attractive, supporting the existence of a $\phi$-$p$ bound state.
The binding energy of the $\phi$-$p$ bound state in the spin 1/2 channel is estimated to lie within the range of 12.8 to 56.1 MeV.

This analysis demonstrates how experimental data on correlation functions, combined with lattice QCD results, can be used to explore potential bound states as an alternative to the invariant mass technique.

\section{Meson-Baryon Interactions with charm: $J/\psi$-$N$ and $\eta_c$-$N$}

\begin{figure*}[b]
    \centering
    \includegraphics[width=7.0cm]{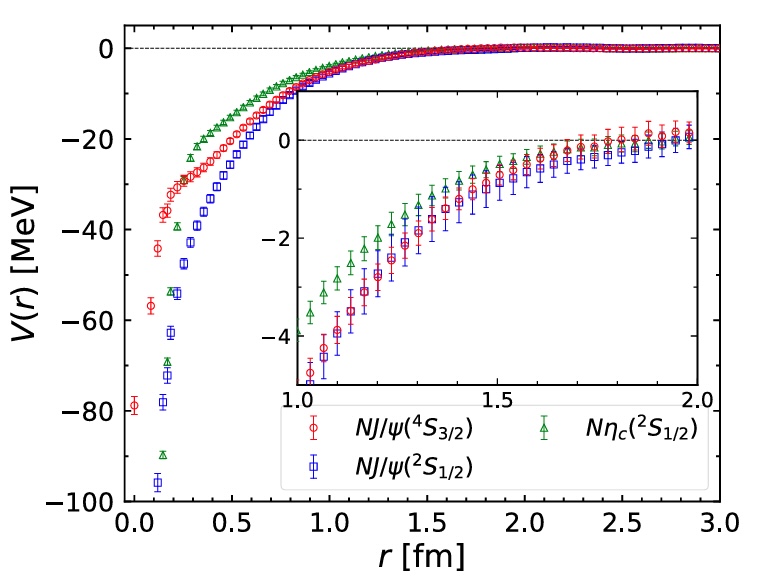}\hspace{0.5cm}
   \includegraphics[width=7.5cm]{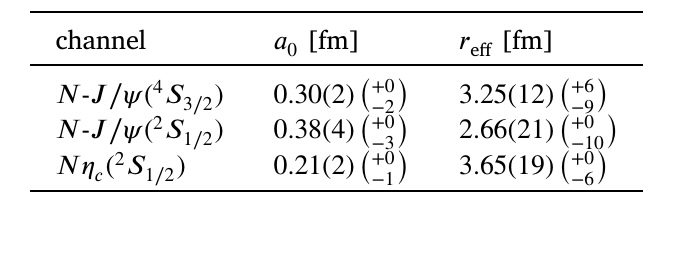} 
   \caption{(Left) The $S$-wave $c\bar{c}$-$N$ potential extracted at $t/a = 14$
   for $J/\psi$-$N$ with $^4S_{3/2}$ and $^2S_{1/2}$ (b), and $\eta_c$-$N$ with $^2S_{1/2}$. 
   (Right) The $c\bar{c}$-$N$ scattering length $a_0$ and effective range $r_{\mathrm{eff}}$ with statistical errors (1st parentheses) and systematic errors (2nd parentheses). 
Taken from \cite{Lyu:2024ttm}.}
    \label{Fig.5}
\end{figure*} 

The charmonium-nucleon interaction is important for several physical reasons.
The first pentaquark candidate \(P_c\) was observed in the \(J/\psi\)-\(p\) spectrum by the LHCb Collaboration \cite{LHCb2015}. A complete understanding of its nature and properties requires a reliable \(c\bar{c}\)-\(N\) interaction, which serves as an essential ingredient for a comprehensive coupled-channel analysis of \(P_c\). The potential existence of charmonium-nucleus bound states (\(c\bar{c}\)-\(A\)) has been extensively discussed (see \cite{Krein2017} and references therein). However, the theoretically predicted binding energies vary significantly due to the considerable uncertainty in the \(c\bar{c}\)-\(N\) interaction.
 Furthermore, the \(c\bar{c}\)-\(N\) interaction contributes to the total \(J/\psi\)-\(N\) cross section, which is relevant for studies of \(J/\psi\) suppression in nuclear collisions, as well as for understanding the intrinsic charm content of the nucleon. Additionally, the \(J/\psi\)-\(N\) scattering length provides insights into the modification of the \(J/\psi\) mass in the nuclear medium.

In Ref.\cite{Lyu:2024ttm}, a realistic study on the low-energy interaction between a nucleon ($N$) and a charmonium ($J/\psi$ and $\eta_c$) based on the 
K-configuration
 with physical charm quark mass was carried out.
The $c\bar{c}$-$N$ potential is found to be attractive at all distances as shown in Fig.\ref{Fig.5} (left panel).
The $S$-wave scattering length $a_0$ and the effective range $r_{\rm eff}$ obtained from the potential are summarized in the right panel of Fig.\ref{Fig.5}. 
The quantitative predictions here could be validated by the  future femtoscopic analysis of $J/\psi$-$p$ at LHC. 
Also, by using the theoretical scattering length, the mass ``reduction" of the  $J/\psi$ meson in nuclear matter under the linear density approximation can be  evaluated as 
\begin{equation}
 \delta m_{_{J/\psi}}  \simeq \frac{2\pi (m_{_N} + m_{_{J/\psi}})}{m_{_N} m_{_{J/\psi}}} a_{_{J/\psi}}^{\text{spin-av}} \rho_{0} = 19(3) \, \text{MeV},
\end{equation}
with $\rho_{0} = 0.17~\mathrm{fm}^{-3}$ being the normal nuclear matter density, and the $J/\psi$-$N$ being the spin-averaged scattering length 
$a_{J/\psi}^{\mathrm{spin-av}} = ({2a_0^{(3/2)} + a_0^{(1/2)}})/3$.
This opens a possibility to form the charmonium-nucleus bound states for sufficiently large nuclei to be explored experimentally \cite{Krein2017}.

\section{Two-pion Exchange (TPE) Paradigm}

\subsection{TPE in nuclear force}
\label{sec:TPE-NN}

\begin{figure*}[t]
    \centering
    \includegraphics[width=14cm]{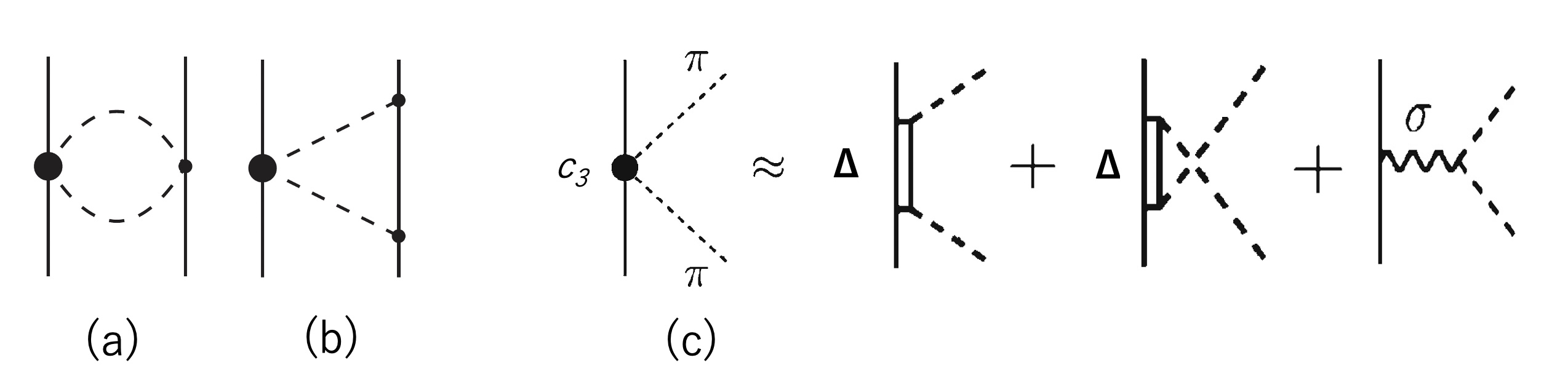}
   \caption{(a,b) The  N$^2$LO  football and triangle diagrams contributing to the TPEP in the NN interaction. The large (small) solid dot is the vertex in the 2nd order (1st order) $\pi N$ Lagrangian  $\widehat{\mathcal{L}}^{(2)}_{\pi N} $  ($\widehat{\mathcal{L}}^{(1)}_{\pi N} $). (b) The physical origin of the 
  $c_3$ vertex in $\widehat{\mathcal{L}}^{(2)}_{\pi N} $  
  from the meson theoretic point of view with  $\Delta$ excitation and the $S$-wave $\pi\pi$ correlation ($\sigma$).}
    \label{Fig6}
\end{figure*}

 In the nuclear force between nucleons, the long-range part of the interaction at distances \( r \sim m_{\pi}^{-1} \) is dominated by the one-pion exchange potential (OPEP), originally proposed by Yukawa~\cite{Yukawa:1935xg}.  
At intermediate distances, \( r \sim (2m_{\pi})^{-1} \), the leading contribution arises from the two-pion exchange potential (TPEP), as demonstrated in the seminal work of Taketani, Machida, and Ohnuma, commonly known as the TMO potential~\cite{TMO:1951}.  
Recent systematic studies based on chiral effective field theory (EFT) have shown that the TPEP first appears at next-to-leading order (NLO). However, the state-independent (i.e., spin- and isospin-independent) central potential \( V_C \) emerges only at next-to-next-to-leading order (N$^2$LO), as illustrated in Fig.~\ref{Fig6}(a,b) (see the reviews~\cite{Machleidt:2011zz,Epelbaum:2008ga} and references therein).  Interestingly, the contribution from the football diagram in Fig.~\ref{Fig6}(a) vanishes, leaving only the triangle diagram in Fig.~\ref{Fig6}(b) as a nonzero contribution.

 Given below are the explicit form of the 
  state-independent TPE NN potential ${V}_{\rm NN;C}^{2\pi}(r) $ at N$^2$LO in the coordinate space \cite{Kaiser:1997mw}:
\begin{eqnarray}
 \left. {V}_{\rm NN;C}^{2\pi}(r) \right|_{\rm N^2LO}^{\rm Fig.6(a)}
 & = & 0, \label{eq:NN-NNLOa} \\
\! \! \!  \left. {V}_{\rm NN;C}^{2\pi}(r) \right|_{\rm N^2LO}^{\rm Fig.6(b)}
 & = &  \hspace{-0.8cm} \frac{3g_A^2}{32\pi^2 f_\pi^4} \frac{e^{-2x}}{r^6}  \left\{  \left( 2c_1 + \frac{3g_A^2}{16M} \right) x^2 (1+x)^2 
 + \frac{g_A^2 x^5}{32M} \right.  \nonumber \label{eq:NN-NNLOb}\\ 
& & \hspace{1.5cm}  + \left. \left( c_3 + \frac{3g_A^2}{16M} \right) (6 + 12x + 10x^2 + 4x^3 + x^4) \right\} , \\ 
& \xrightarrow[M\rightarrow \infty]{} &  
 \frac{3g_A^2}{32\pi^2 f_\pi^4} \frac{e^{-2x}}{r^6}    \left\{ 2c_1  x^2 (1+x)^2  + c_3 (6 + 12x + 10x^2 + 4x^3 + x^4) \right\} , \nonumber \\
 & & \\
& \xrightarrow[M\rightarrow \infty,\ x \ll 1]{} & 
 \frac{9g_A^2 c_3}{16\pi^2 f_\pi^4}\cdot \frac{1}{r^6} ,   \label{vanderW} \\
& \xrightarrow[M\rightarrow \infty,\ x \gg 1]{} & 
 \frac{3g_A^2 c_3 m_{\pi}^4 }{32\pi^2 f_\pi^4} \cdot \frac{e^{-2m_{\pi}r} }{r^2}   ,
\label{yukawa2}
\end{eqnarray}
with $M$ being the nucleon mass and $x= m_{\pi} r$.  Here $g_A/f_{\pi}$ originates from the 
the 1st-order $\pi N$ Lagrangian $\widehat{\mathcal{L}}^{(1)}_{\pi N}$
in the heavy baryon formalism, while
 $c_{1,3}$  are the low-energy constants in the 2nd-order chiral Lagrangian $\widehat{\mathcal{L}}^{(2)}_{\pi N}$ \cite{Machleidt:2011zz}:
\begin{eqnarray}
\widehat{\mathcal{L}}^{(1)}_{\pi N} &=& \bar{N} \left\{ i \partial_0 
- \frac{1}{4 f_\pi^2} \boldsymbol{\tau} \cdot \left( \boldsymbol{\pi} \times \partial_0 \boldsymbol{\pi} \right) 
- \frac{g_A}{2 f_\pi} \boldsymbol{\tau} \cdot \left( \vec{\sigma} \cdot \vec{\nabla} \right) \boldsymbol{\pi} + \cdots
\right\} N,
\label{eq:piN-1}\\
\widehat{\mathcal{L}}^{(2)}_{\pi N} &=& 
\bar{N} \Bigg[ 
4c_1 m_\pi^2 
- \frac{2c_1}{f_\pi^2} m_\pi^2 \, \boldsymbol{\pi}^2 
+ \left( c_2 - \frac{g_A^2}{8M} \right) \frac{1}{f_\pi^2} (\partial_0 \boldsymbol{\pi} \cdot \partial_0 \boldsymbol{\pi}) 
+ \frac{c_3}{f_\pi^2} (\partial_\mu \boldsymbol{\pi} \cdot \partial^\mu \boldsymbol{\pi}) \nonumber\\
& & \hspace{1cm} - \left( c_4 + \frac{1}{4M} \right) \frac{1}{2f_\pi^2} 
\, \epsilon^{ijk} \epsilon^{abc} \sigma^i \tau^a 
(\partial^j \pi^b)(\partial^k \pi^c) 
\Bigg] N + \cdots ,
\label{eq:piN-2}
\end{eqnarray}
where  $\cdots$ indicates the terms with 
three or more pions.
The  empirical values of $c_{1-4}$ from 
Table 1 of \cite{Machleidt:2020vzm} read,  
 $c_{1} = -1.10(3)$,
  $c_{2} = 3.57(4)$,
 $c_{3} = -5.54(6)$,
 and  $c_{4} = 4.17(4)$,
  in the unit of ${\rm GeV}^{-1} $.

In the static limit, where the nucleon mass \( M \) is large, the dominant contribution to TPE arises from the vertex proportional to \( c_3 \) in Eq.~(\ref{eq:piN-2}), combined with the standard vertex proportional to \( g_A \) in Eq.~(\ref{eq:piN-1}).  
Furthermore, Eq.~(\ref{vanderW}) shows that in the limit of a massless pion, the static TPEP takes the form of a van der Waals potential, while Eq.~(\ref{yukawa2}) demonstrates that in the asymptotic region, the TPEP behaves proportionally to the square of the OPEP.

It is noteworthy that the low-energy constant \( c_3 \) plays a crucial role. Microscopically, it is related to \(\pi N\) scattering involving \(\Delta\) excitation and to the $S$-wave \(\pi\pi\) correlation (the $\sigma$ meson)~\cite{Kaiser:1998wa,Machleidt:2020vzm}, as illustrated in Fig.~\ref{Fig6}(c).  
Since both \( 2c_1 + \frac{3g_A^2}{16M} \) and \( c_3 + \frac{3g_A^2}{16M} \) are negative, the state-independent part of the TPEP, given by Eq.~(\ref{eq:NN-NNLOb}), becomes attractive at intermediate and long distances.

\subsection{TPE between flavor-singlet hadrons}
\label{sec:TPE-FSFS}

 \begin{figure*}[t]
\hspace{1cm}
    \centering
    \includegraphics[width=9.5cm]{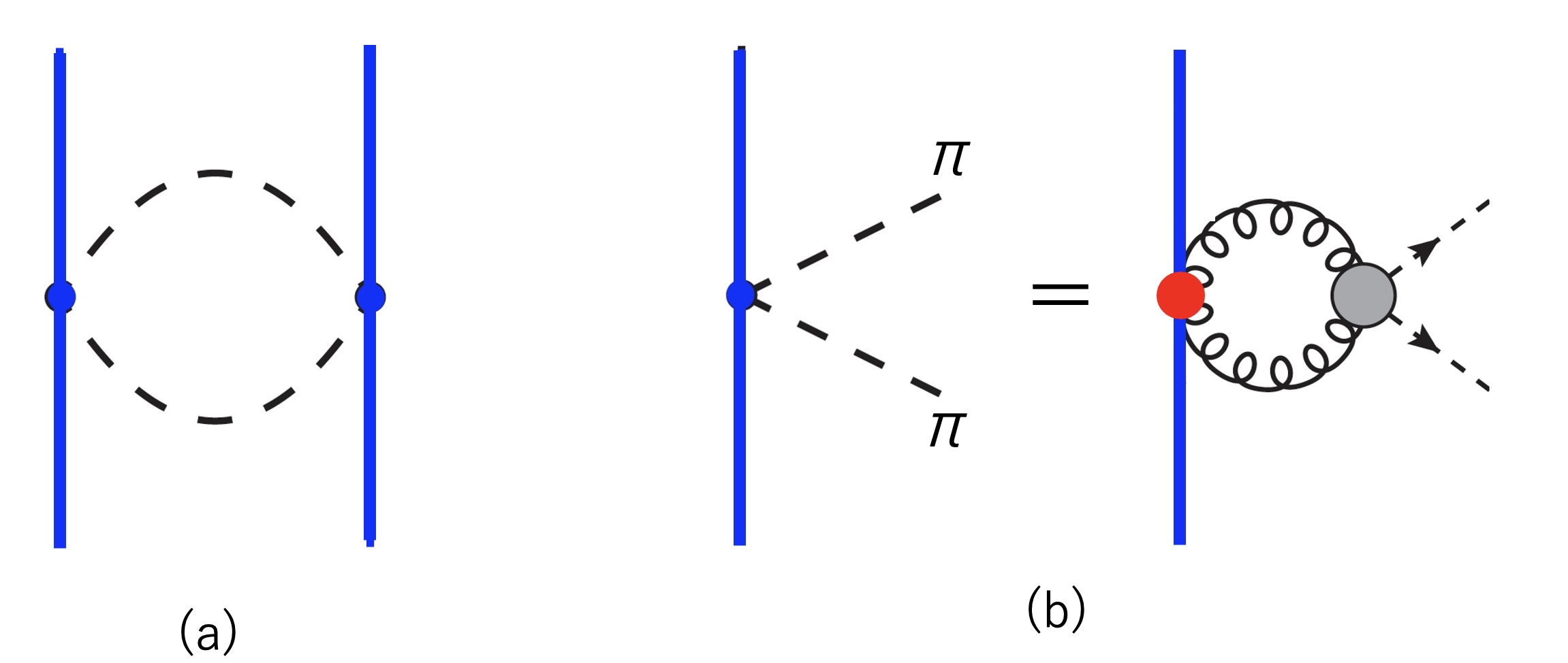}
   \caption{(a) The long-range interaction between the two flavor-singlet hadrons (the blue lines) through the two-pion exchange.  (b) The microscopic origin of the coupling of a
   flavor singlet hadron and the two pions (the solid blue dot) through
    the gluon cloud \cite{Brambilla:2015rqa}.}
    \label{Fig.7}
\end{figure*} 

If both nucleons in Sec.~\ref{sec:TPE-NN} are replaced by flavor-singlet hadrons, OPEP is no longer allowed. In this case, the longest-range interaction must arise from TPEP. A characteristic example is the interaction between two heavy quarkonia, such as bottomonium-bottomonium or charmonium-charmonium systems, as illustrated by the football diagram in Fig.~\ref{Fig.7}(a). The gluons surrounding each \( Q\bar{Q} \) pair couple to flavor-singlet two-pion states [Fig.~\ref{Fig.7}(b)], which can be exchanged between the dipoles to generate the TPEP. This type of interaction was first investigated by Peskin and Bhanot~\cite{Bhanot:1979vb}, and was later clarified using dispersion relations~\cite{Fujii:1999xn} and chiral EFT~\cite{Brambilla:2015rqa}.
An explicit expression for the interaction between heavy quarkonia was derived in Ref.~\cite{Brambilla:2015rqa} and is given below:
 \begin{eqnarray}
\left. V^{2\pi}_{Q\bar{Q}-Q\bar{Q};C}(r) \right|_{\rm Fig.7(a)}
& = &  -\frac{3\pi \beta^2 m_{\pi}^2}{8b^2 r^5} 
 \sum_{n=1,2} g_n(x) K_n (2x) , 
 \label{full-Q}\\
& \xrightarrow[x \gg 1]{} & -\frac{3(3+\kappa_2)^2 \pi^{3/2} \beta^2 m_{\pi}^{9/2} }{4b^2}  \cdot \frac{e^{-2m_{\pi}r}}{r^{5/2}}.
\label{asymp}
\end{eqnarray}
 Here $x=m_{\pi}r$,
 $g_1(x)=  4(\kappa_2 + 3)^2 x^3
+ (3\kappa_1^2 + 43\kappa_2^2 + 14\kappa_1 \kappa_2) x $, $g_2(x) = 2(2(\kappa_2 + 3)(\kappa_1 + 5\kappa_2)x^2 
+ 3\kappa_1^2 + 43\kappa_2^2 + 14\kappa_1 \kappa_2)$ and
 $K_n(y)$ is the modified Bessel functions of the
second kind. 
Also, $ b = 11 - \frac{2}{3} N_f $,  
$\beta$ denotes the color-electric polarizability of the quarkonium ground state, and \( \kappa_{1,2} \) are parameters related to the coupling of two pions to the chromoelectric field strength.  
Note that the asymptotic form 
Eq.(\ref{asymp}) significantly underestimates the full result Eq.(\ref{full-Q}) at distances \( r \sim (2m_{\pi})^{-1} \).  

Eq.~(\ref{full-Q}) provides a suitable starting point for fitting lattice QCD data on \( J/\psi\text{--}J/\psi \) and \( \eta_c\text{--}\eta_c \) interactions within the HAL QCD method.  
Furthermore, extending this formula to heavy baryons would be valuable for investigating the long-range behavior of the \( \Omega_{sss}\text{--}\Omega_{sss} \) and \( \Omega_{ccc}\text{--}\Omega_{ccc} \) interactions, for which lattice data have already been obtained by the HAL QCD Collaboration~\cite{Gongyo:2017fjb,Lyu:2021qsh}.

\subsection{TPE between nucleon and flavor-singlet hadron}
\label{sec:TPE-FSN}

\begin{figure*}[t]
    \centering
    \includegraphics[width=11cm]{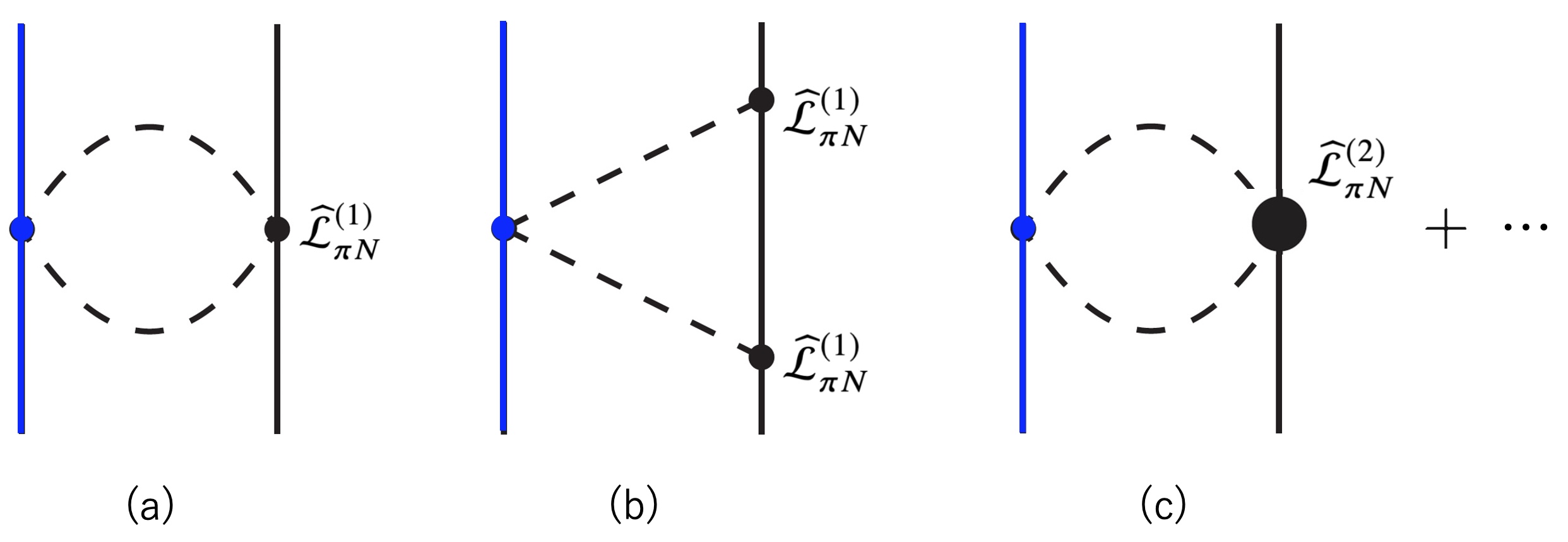}
  \caption{(a) The N$^3$LO football TPE diagram between a flavor-singlet hadron (the blue line) and the nucleon through the two-pion exchange. The blue solid dot is given by 
   Fig.\ref{Fig.7}(b), while the small 
   black solid dot is from the $\pi \pi NN$ seagull term 
   in $\widehat{\mathcal{L}}^{(1)}_{\pi N} $.  This contribution turns out to be zero due to isospin structure of the vertices \cite{TarrusCastella:2018php}.  
   (b) The N$^3$LO triangle TPE diagram with the small   black solid dot from the $\pi NN$ term
    proportional to $g_A$ in $\widehat{\mathcal{L}}^{(1)}_{\pi N} $.
    (c) The N$^4$LO contributions including the football TPE diagram with the large black solid dot from the terms
    in $\widehat{\mathcal{L}}^{(2)}_{\pi N} $.}
    \label{Fig.8}
\end{figure*}

If only one of the interacting hadrons is flavor-singlet,  it is an intermediate situation between the cases  discussed in Sec.\ref{sec:TPE-NN} and 
Sec.\ref{sec:TPE-FSFS}. 
Since the flavor-singlet hadron is involved, OPEP is still inhibited and the possible long-range interaction should come from TPEP
first appearing in the N$^3$LO (Fig.\ref{Fig.8}(a,b)).  These diagrams  in the chiral EFT approach
 have been  evaluated  as \cite{TarrusCastella:2018php} 
\begin{eqnarray}
\left. V^{2\pi}_{Q\bar{Q}-N;C} (r) \right|_{\rm N^3LO}^{{\rm Fig.8(a)}} 
& = & 0,\\
\left. V^{2\pi}_{Q\bar{Q}-N;C} (r) \right|_{\rm N^3LO}^{{\rm Fig.8(b)}} 
& = & 
  \frac{3g_A^2}{128\pi^2 f_{\pi}^2 r^6} e^{-2x } 
\Big\{ c_{{di}} g_d(x)
+ c_m g_m(x)
\Big\},  
\label{full-2}\\
& \xrightarrow[x \gg 1]{} &
 \frac{3g_A^2 (c_{{di}} + c_m) m_\pi^4 }{128\pi^2 f_{\pi}^2 } \frac{e^{-2m_\pi r}}{r^2},
 \label{asymp-2}
\end{eqnarray}
where $x=m_{\pi}r$, $g_d(x)= 6 + x (2 + x)(6 + x (2 + x))$ and  $g_m(x) = x^2 (1 + x)^2$. Also, 
$c_{d0} = -\frac{4\pi^2 \beta}{b} \kappa_1$, 
$c_{di} = -\frac{4\pi^2 \beta}{b} \kappa_2$, 
and $c_m = -\frac{12\pi^2 \beta}{b}$.
 Note that the asymptotic form Eq.(\ref{asymp-2}) significantly underestimates 
 Eq.(\ref{full-2}) at distances $r \sim (2m_{\pi})^{-1}$.

In Fig.~\ref{Fig.8}(c), the contributions at N$^4$LO are shown, highlighting the 
football diagram that includes the $\pi\pi NN$ vertex from $\widehat{\mathcal{L}}^{(2)}_{\pi N}$.
Although this diagram is formally of higher order compared to the triangle diagram 
in Fig.~\ref{Fig.8}(b), it may lead to an enhancement, similar to the case of two-pion exchange (TPE) in the $NN$ interaction 
$\left. {V}_{\rm NN;C}^{2\pi}(r) \right|_{\rm N^2LO}^{\rm Fig.6(b)}$ occurring through the $c_3$-term, driven by the $\Delta$ excitation. Further details of this analysis will be reported elsewhere.

\subsection{Signatures of TPE in $\phi$-N, $J/\psi$-N and $\eta_c$-N interactions}
\label{sec:TPE-lattice}

To check if the TPE 
 between a flavor-singlet hadron and a nucleon
  discussed in Sec.\ref{sec:TPE-FSN} arises in reality, 
  let consider the lattice QCD potentials $V_{\rm lat}$
  for  $\phi$-N, $J/\psi$-N and $\eta_c$-N interactions
 given in  \cite{Lyu:2022imf,Lyu:2024ttm} and 
   introduce ``spatial effective energy" as a function of $r$:
\begin{eqnarray}
E_{\text{eff}}(r) = -\frac{\ln[-V_{\rm lat}(r) r^\nu / \alpha]}{r}.
\end{eqnarray}
 If two-pion exchange (TPE) dominates at distances \( r > (2m_\pi)^{-1} \),  \( E_{\text{eff}}(r) \) should approach the known value \( 2m_\pi = 292.8\,\text{MeV} \) of the K-configuration.

\begin{figure*}[t]
    \centering
    \includegraphics[width=7cm]{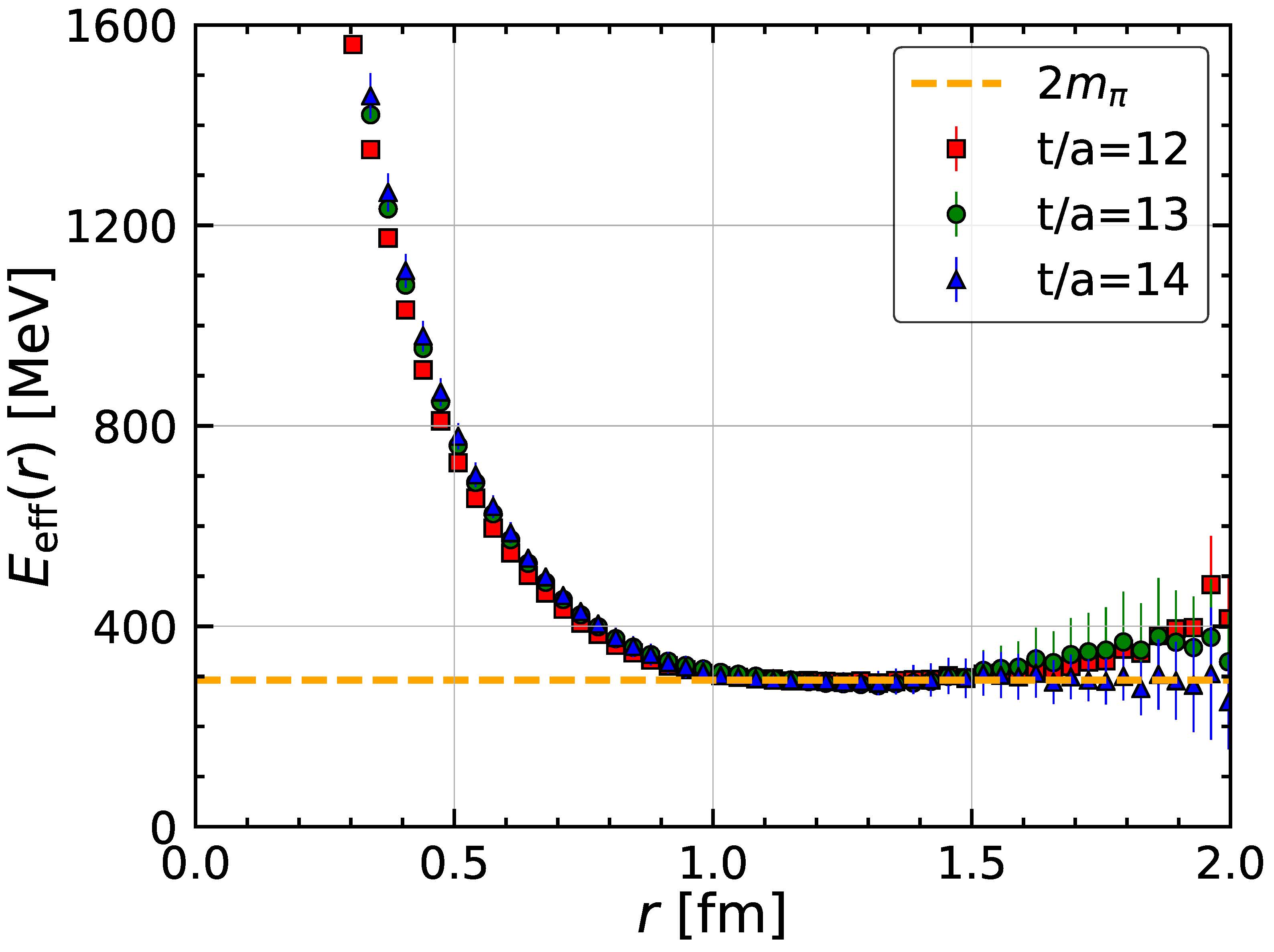}\hspace{0.5cm}
   \includegraphics[width=7cm]{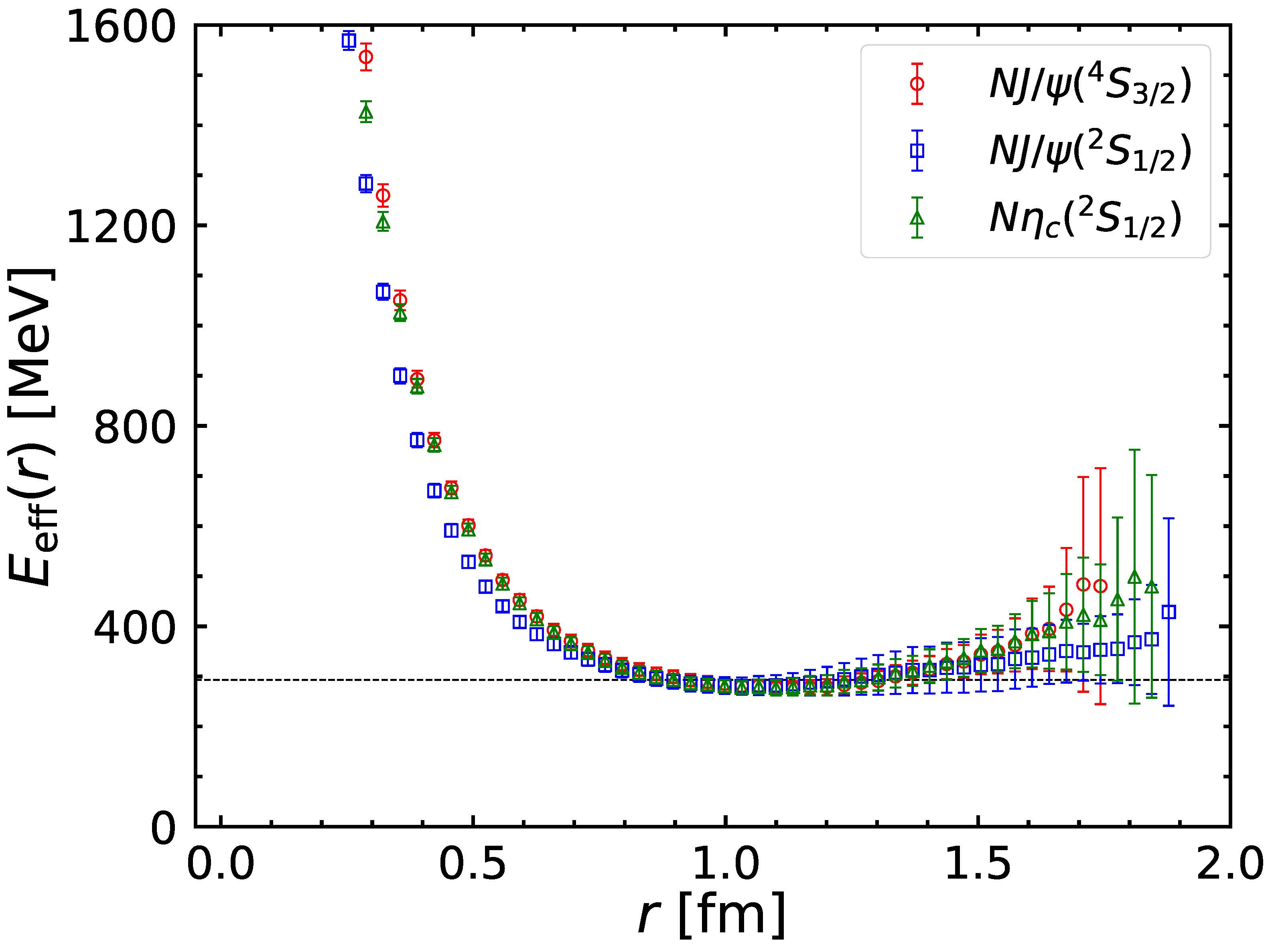} 
   \caption{(Left) 
   The spatial effective energy \( E_{\text{eff}}(r) \) as a function of separation \( r \) at Euclidean time \( t/a = 12 \) (red squares), 13 (green circles), and 14 (blue triangles). The orange dashed line corresponds to \( 2m_\pi \) with the known value of the pion mass \( m_\pi = 146.4 \) MeV in the K-configuration \cite{Lyu:2022imf}.
    (Right)  \( E_{\text{eff}}(r) \) for 
     two spin states (\( J = 1/2 \) and \( 3/2 \)) of the \( J/\psi \)\nobreakdash-\(N \) system, and the \( J = 1/2 \) state of the \(\eta\)\nobreakdash-\(N \) system at \( t/a = 14 \). The black dotted line corresponds to \( 2m_\pi = 292.8 \) MeV
     (unpublished results based on~\cite{Lyu:2024ttm}).
   }
    \label{Fig.9}
\end{figure*}

 In the present analysis, we adopt a simple choice of \( \nu = 2 \), motivated by Eq.~(\ref{asymp-2}).  
(This could be replaced by \( \nu = 5/2 \), as discussed at the end of Sec.~\ref{sec:TPE-FSN}.)  
 The constant $\alpha$ is determined by
  fitting the lattice data in the range 
  $0.9\ {\rm fm}  < r < 1.8\ {\rm fm}$ with $t/a=14$.
   The left panel of Fig.~\ref{Fig.9} shows \( E_{\text{eff}}(r) \) for the \(\phi\)\nobreakdash-\(N\) interaction at \( t/a = 12, 13, 14 \)~\cite{Lyu:2022imf}, while the right panel presents results for two different spin states (\( J = 1/2 \) and \( 3/2 \)) of the \( J/\psi \)\nobreakdash-\(N \) system, as well as the \( J = 1/2 \) state of the \(\eta_c\)\nobreakdash-\(N \) system at \( t/a = 14 \) (unpublished results based on~\cite{Lyu:2024ttm}).
   
     In all cases, 
We observe that $E_{\text{eff}}(r)$ reaches a plateau at $2m_\pi = 292.8 \, \text{MeV}$ for $r > 1.0 \, \text{fm}$ within the statistical uncertainty,
 suggesting that the long-range part 
 of the interaction is dominated by the TPEP.
 To firmly establish this conclusion, further improvements are needed—both in reducing statistical uncertainties in lattice QCD calculations and in advancing the theoretical analysis of the football diagram in Fig.~\ref{Fig.8}(c) within the framework of chiral EFT, as suggested in Sec.~\ref{sec:TPE-FSN}.

\section{Summary}

In this article, we summarize recent developments in the study of hadron-hadron interactions using lattice QCD near the physical pion mass ($m_{\pi} \simeq 146$~MeV), based on the HAL QCD method and its connection to experimental data. In particular, we focus on several key interaction channels:
\begin{itemize}
    \item \textbf{Baryon-Baryon Interactions:} The $\Lambda\Lambda$--$\Xi N$ coupled-channel system reveals short-range channel mixing and a mid-to-long-range attractive $\Xi N$ potential, consistent with hypernuclear spectroscopic data from J-PARC as well as femtoscopic correlation measurements at RHIC and the LHC.

    \item \textbf{Meson-Meson Interactions:} The $D^\ast$--$D$ interaction provides insights into the structure of the doubly charmed tetraquark state $T_{cc}$, supporting its interpretation as a possible shallow and narrow quasi-bound state observed at the LHC.

    \item \textbf{Meson-Baryon Interactions:} The $\phi$--$N$, $J/\psi$--$N$, and $\eta_c$--$N$ potentials exhibit attractive behavior across all distances. This analysis paves the way for exploring interactions between nucleons and flavor-singlet hadrons by combining lattice QCD results with experimental data.
\end{itemize}

Furthermore, we examine the two-pion exchange (TPE) mechanism, which governs the long-range behavior of interactions between flavor-singlet hadrons, as well as between nucleons and flavor-singlet hadrons. In the latter case, signatures of TPE have been identified in the lattice QCD potentials. We also highlight the possible importance of the $\pi\pi NN$ seagull term from the second-order chiral Lagrangian, which is known to play a dominant role in TPE contributions to the $NN$ potential.

The results presented in this article demonstrate the HAL QCD method as a robust and quantitative tool for probing nonperturbative QCD dynamics, effectively bridging lattice predictions with experimental observations.

\section*{Acknowledgements}

The author gratefully acknowledges the invaluable contributions and insightful discussions with members of the HAL QCD Collaboration. In particular, I would like to thank Yan Lyu and Takumi Doi for the fruitful discussions on the two-pion exchange mechanism between hadrons.
 I would also like to thank Lingxiao Wang
 for the illuminating discussion on the reconstruction of the non-local potential by the deep neural network. This work was partly supported by Japan Science and Technology Agency (JST) as part of Adopting Sustainable Partnerships for Innovative Research Ecosystem (ASPIRE), Grant Number JPMJAP2318.


\newpage

\begin{thebibliography}{99}

\bibitem{Fugaku-conf}
T.~Aoyama, T. M. Doi, T. Doi, E. Itou, Yan Lyu,
 K. Murakami and T. Sugiura [HAL QCD Collaboration], Phys. Rev. D \textbf{110} (2024) 094502.

\bibitem{Baym2018}
G.~Baym, T.~Hatsuda, T.~Kojo, P.~D.~Powell, Y.~Song and T.~Takatsuka,
Rept. Prog. Phys. \textbf{81} (2018) 
 056902.

\bibitem{Hayano2008}
R.~S.~Hayano and T.~Hatsuda, Rev. Mod. Phys. \textbf{82} (2010) 2949.

\bibitem{HAL1}
N. Ishii, S. Aoki and T. Hatsuda, Phys. Rev. Lett. {\bf 99} (2007) 022001.
\bibitem{HAL2}
S. Aoki, T. Hatsuda and N. Ishii, Prog. Theor. Phys. {\bf 123} (2010) 89.

\bibitem{HAL3}
N. Ishii, S. Aoki, T. Doi, T. Hatsuda, Y. Ikeda, 
T. Inoue,  K. Murano, H. Nemura,  K. Sasaki [HAL QCD Collaboration],  Phys. Lett. B {\bf 712} (2012) 437.

\bibitem{HAL4}
S. Aoki and T. Doi, in {\it Handbook of Nuclear Physics} (eds. I. Tanihata, H. Toki and T. Kajino), pp.1787-1817 (2023)  [arXiv:2402.11759 [hep-lat]]. 

\bibitem{Reduction}
W.~Zimmermann,
MPI-PAE/PTh-61/87 (1987), published in
{\it Wandering in the Fields: Festschrift for Professor Kazuhiko Nishijima on the Occasion of His Sixtieth Birthday}
(eds. K. Kawarabayashi and A. Ukawa), World-Scientific, Singapore, 1987.

\bibitem{FVM1986}
M. L\"{u}scher, Commun. Math. Phys. {\bf 104} (1986) 177; ibid., {\bf 105} (1986) 153; Nucl. Phys. B {\bf 354} (1991) 531.


\bibitem{HALReview}
S.~Aoki and T.~Doi,
Front. in Phys. \textbf{8} (2020) 307.

\bibitem{K-conf} K.I. Ishikawa \textit{et al.}  [PACS Collaboration],
PoS {\bf LATTICE2015} (2016) 075.



\bibitem{Wang2025}
L.~Wang, T.~Doi, T.~Hatsuda and Y.~Lyu,
PoS \textbf{LATTICE2024} (2025) 076.


\bibitem{Sasaki2020}
K. Sasaki, S. Aoki, T. Doi, S. Gongyo, T. Hatsuda, Y. Ikeda, T. Inoue, T. Iritani, N. Ishii, K. Murano, T. Miyamoto, [HAL QCD Collaboration],
Nucl. Phys. A \textbf{998} (2020) 121737.

\bibitem{Kamiya2022}
Y.~Kamiya, K.~Sasaki, T.~Fukui, T.~Hyodo, K.~Morita, K.~Ogata, A.~Ohnishi and T.~Hatsuda,
Phys. Rev. C \textbf{105} (2022) 014915.

\bibitem{Isaka2024}
M.~Isaka, T.~Tada, M.~Kimura and Y.~Yamamoto,
Phys. Rev. C \textbf{109} (2024) 044317.

\bibitem{ALICE2020}
S. Acharya   \textit{et al.}
[ALICE Collaboration],
Nature \textbf{588} (2020) 232.

\bibitem{LHCb2022}
R.Aaij \textit{et al.} [LHCb Collaboration], Nature Phys. {\bf 18} (2022) 751; Nature Commun. {\bf 13}(2022) 3351.


\bibitem{Lyu2024}
Y.~Lyu, S.~Aoki, T.~Doi, T.~Hatsuda, Y.~Ikeda and J.~Meng,
Phys. Rev. Lett. \textbf{131} (2023) 161901.

\bibitem{ALICE2021}
S. Acharya \textit{et al.} [ALICE Collaboration],
Phys. Rev. Lett. {\bf 127} (2021) 172301.

\bibitem{Lyu:2022imf}
Y.~Lyu, T.~Doi, T.~Hatsuda, Y.~Ikeda, J.~Meng, K.~Sasaki and T.~Sugiura,
Phys. Rev. D \textbf{106} (2022) 074507.


\bibitem{Chizzali2024}
E.~Chizzali, Y.~Kamiya, R.~Del Grande, T.~Doi, L.~Fabbietti, T.~Hatsuda and Y.~Lyu,
Phys. Lett. B {\bf 848} (2024) 138358.


\bibitem{Iritani2019}
T. Iritani, S. Aoki, T. Doi, F. Etminan, 
S. Gongyo,  T. Hatsuda,  Y. Ikeda, T. Inoue,  N. Ishii, 
T. Miyamoto,  K. Sasaki [HAL QCD Collaboration],
Phys. Lett. B {\bf 792} (2019) 284.


\bibitem{LHCb2015}
R. Aaij \textit{et al.} [LHCb Collaboration],
 Phys. Rev. Lett. {\bf 115} (2015) 072001; ibid.
 {\bf 122} (2019) 222001.

\bibitem{Krein2017}
G.~Krein, A.~W.~Thomas and K.~Tsushima,
Prog. Part. Nucl. Phys. \textbf{100} (2018) 161.


\bibitem{Lyu:2024ttm}
Y.~Lyu, T.~Doi, T.~Hatsuda and T.~Sugiura,
Phys. Lett. B \textbf{860} (2025) 139178.


\bibitem{Yukawa:1935xg}
H.~Yukawa,
Proc. Phys. Math. Soc. Jap. \textbf{17} (1935) 48.

\bibitem{TMO:1951}
M. Taketani, S. Machida and Onuma,
 Prog. Theor. Phys. {\bf 6} (1951) 638(L);
 {\it ibid.} {\bf 7} (1952) 45.

\bibitem{Machleidt:2011zz}
R.~Machleidt and D.~R.~Entem,
Phys. Rept. \textbf{503} (2011) 1.

\bibitem{Epelbaum:2008ga}
E.~Epelbaum, H.~W.~Hammer and U.~G.~Meissner,
Rev. Mod. Phys. \textbf{81} (2009) 1773.

\bibitem{Kaiser:1997mw}
N.~Kaiser, R.~Brockmann and W.~Weise,
Nucl. Phys. A \textbf{625} (1997) 758.

\bibitem{Kaiser:1998wa}
N.~Kaiser, S.~Gerstendorfer and W.~Weise,
Nucl. Phys. A \textbf{637} (1998) 395.

\bibitem{Machleidt:2020vzm}
R.~Machleidt and F.~Sammarruca,
Eur. Phys. J. A \textbf{56} (2020) 95.

\bibitem{Bhanot:1979vb}
G.~Bhanot and M.~E.~Peskin,
Nucl. Phys. B \textbf{156} (1979) 391.

\bibitem{Fujii:1999xn}
H.~Fujii and D.~Kharzeev,
Phys. Rev. D \textbf{60} (1999) 114039.

\bibitem{Brambilla:2015rqa}
N.~Brambilla, G.~Krein, J.~Tarr\'us Castell\`a and A.~Vairo,
Phys. Rev. D \textbf{93} (2016) 054002.

\bibitem{Gongyo:2017fjb}
S.~Gongyo, K.~Sasaki, S.~Aoki, T.~Doi, T.~Hatsuda, Y.~Ikeda, T.~Inoue, T.~Iritani, N.~Ishii and T.~Miyamoto, \textit{et al.}
Phys. Rev. Lett. \textbf{120} (2018) 212001.


\bibitem{Lyu:2021qsh}
Y.~Lyu, H.~Tong, T.~Sugiura, S.~Aoki, T.~Doi, T.~Hatsuda, J.~Meng and T.~Miyamoto,
Phys. Rev. Lett. \textbf{127} (2021) 072003.

\bibitem{TarrusCastella:2018php}
J.~Tarr\'us Castell\`a and G.~Krein,
Phys. Rev. D \textbf{98} (2018) 014029.






\end{thebibliography}
\end{document}